\documentclass[preprint2]{aastex}
\usepackage{natbib}
\usepackage[dvips]{color}
\newif\ifAMStwofonts
\AMStwofontstrue

\def\vu{\bf u}

\def\gsim{~\rlap{$>$}{\lower 1.0ex\hbox{$\sim$}}}

\def\simpropto{\lower.2ex\hbox{$\; \buildrel \propto \over \sim \;$}}
\def\ltsim{\lower.5ex\hbox{$\; \buildrel < \over \sim \;$}}
\def\gtsim{\lower.5ex\hbox{$\; \buildrel > \over \sim \;$}}
\def\ltsim{\lower.5ex\hbox{$\; \buildrel < \over \sim \;$}}
\def\gtsim{\lower.5ex\hbox{$\; \buildrel > \over \sim \;$}}

\def\kms{\mbox{km\,s$^{-1}$}}

\def\kms{\ {\rm km\,s^{-1}}}
\def\hmpc{\ {\rm h^{-1}Mpc}}

\def\ln{{\rm ln}}

\def\pmb#1{\setbox0=\hbox{#1}%
\kern-.025em\copy0\kern-\wd0
\kern.05em\copy0\kern-\wd0
\kern-.025em\raise.0433em\box0}

\def\simlt{\lower.5ex\hbox{$\; \buildrel < \over \sim \;$}}
\def\simgt{\lower.5ex\hbox{$\; \buildrel > \over \sim \;$}}

\newcommand{\beq}{\begin{equation}}
\newcommand{\eeq}{\end{equation}}
\def\beqa{\begin{eqnarray}}
\def\eeqa{\end{eqnarray}}
\def\fixit#1{}
\def\hmpc{h^{-1}\,{\rm Mpc}}

\shorttitle{Growth rate from z-surveys }
\shortauthors{Nusser, Branchini \& Davis}
\begin{document}

\title{A new method for the determination of the growth rate from  galaxy redshift surveys}
\author{Adi Nusser\altaffilmark{1}, Enzo Branchini\altaffilmark{2}, and Marc Davis\altaffilmark{3}}
\affil{$^1$Physics Department and the Asher Space Science Institute-Technion, Haifa 32000, Israel}
\affil{$^2$Department of Physics, Universit\`a Roma Tre, Via della Vasca Navale 84, 00146, Rome, Italy}
\affil{INFN Sezione di Roma 3, Via della Vasca Navale 84, 00146, Rome, Italy\\ INAF, Osservatorio Astronomico di Brera, Milano, Italy}
\affil{$^3$ Departments of Astronomy \& Physics, University of California, Berkeley, CA. 94720}
\altaffiltext{1}{E-mail: adi@physics.technion.ac.il}
\altaffiltext{2}{E-mail: branchin@fis.uniroma3.it}
\altaffiltext{3}{E-mail: mdavis@berkeley.edu}


\small
\begin{abstract}

Given a redshift survey of galaxies with measurements of apparent magnitudes, we present a novel method for measuring the growth rate $f(\Omega)$
of cosmological linear perturbations. We use the galaxy distribution within the survey to solve for the peculiar velocity field which depends  in linear perturbation theory on  $\beta=f(\Omega)/b$,
where $b$ is the bias factor of the galaxy distribution. 
The recovered line-of-sight peculiar velocities are subtracted from the redshifts to derive the distances, which thus allows an estimate of the absolute magnitude of each galaxy.  A constraint on $\beta$ is then found by minimizing the spread of the estimated magnitudes from their distribution function.  We apply the method to the all sky $K = 11.25$ Two-MASS Redhsift Survey (2MRS) and derive $\beta=0.35\pm 0.1$ at $z\sim 0$, remarkably consistent with our previous estimate from the velocity-velocity comparison. The method could easily be applied to subvolumes extracted from the SDSS survey to derive the growth rate at  $z \sim 0.1$. Further, it should also be applicable to ongoing and future spectroscopic redshift surveys to trace the evolution of $f(\Omega)$ to $z\sim1$.
Constraints obtained from this method are entirely independent from those obtained from the two-dimensional distortion of $\xi(s)$ and provide an important check on $f(\Omega)$, as alternative gravity models predict observable differences.

 \end{abstract}


\section{Introduction}

Large scale density perturbations in the Universe are gravitationally unstable and grow via linear theory.
 The growing mode of  large scale density perturbations, $D(a)$,   
  is characterized by the more observationally relevant   growth rate  
  \begin{equation}
  f(\Omega)=\frac{d \ln D }{ d \ln a}\; ,
  \end{equation}  where $a$ is the scale factor of the Universe
  and   $\Omega $ is the matter density parameter.  It is found that  the
growth index $\gamma=d \ln f / d \ln \Omega $
 is very well approximated by $\gamma=0.55+0.05[1+w(z=1)]$ \citep{Lind05} for 
 a cosmological background dominated by dark energy with an equation of state, $P=w \rho c^2$.
 The growth rate is not only sensitive to the background cosmology, but 
 also to the theory of gravitation invoked  as the driver for 
  structure formation.  Geometric  $R^n$ \citep[e.g.][]{Gann09},
and   Dvali-Gabadadze-Porrati (DGP)  \citep{DGP,Wei}  gravity models give substantially 
different behaviors of $f$.  Models with dark sector 
long range forces, in addition to gravity, even introduce a scale
 dependence into $f$ \citep[e.g.][]{knp10}.

Here we present a new method for constraining $f(\Omega)$ from redshift surveys of galaxies
with measured apparent magnitudes.  
Redshifts of galaxies systematically differ from the actual   distances
by the line-of-sight components of their  peculiar velocities.   Hence,  the directly {\it measurable}  intrinsic luminosities 
or absolute magnitudes of galaxies 
inferred from the observed flux using redshifts rather than distances will
show  larger spread than the true values. 

Gravitational instability theory allows a prediction of  the peculiar velocity field from
observed galaxy distribution given   $f(\Omega)$ and the biasing relation between galaxies and mass.
The method uses this predicted velocity field to get distances for deriving true absolute magnitudes.
 Constraints on $f$ can then be derived by  minimizing the scatter of  the estimated absolute
 magnitudes  from a reference distribution. 
Since the galaxy distribution could be biased relative to the 
mass density field,  the constraints on $f$ are degenerate with the assumed 
biasing. Adopting linear biasing,  $\delta_g=b \delta_m$, between 
the galaxy number density contrast, $\delta_g$, and the mass density contrast, $\delta_m$, yields 
constraints on $\beta=f/b$ that are independent from those obtained from
the apparent anisotropy in the observed galaxy clustering \citep{k88}.
 
In \S2  we describe  the method in detail, presenting  general expressions and deriving the 
relevant approximations. In  \S3 we offer  general analytic assessments  of the method. 
In \S4 we apply the method to the 2MRS  of galaxies  limited to magnitude $K=11.25$ \citep{fhuch}. 
We conclude in \S5 with a general discussion of the results and of the prospects for the 
application of the method to future data.

\section{The method}

This section is largely based on \cite{nbd11a}.
We are given a flux limited survey of galaxies with observed apparent magnitudes $m<m_l$,
angular positions, and  redshifts $cz$ (in $ \kms$). 
Let $r$ (also in $\kms$) be the  
luminosity distance to a galaxy in the sample. For simplicity of notation and description we assume here that the distance and spatial 
extent of the survey are small so that $r$ is well approximate by the physical distance. Therefore, $cz=r+v$ where
$v=\hat r \cdot \bf u$  is the line-of-sight component of 
 three dimensional peculiar velocity $\vu$ of the galaxy. 
 The results can  readily be extended to the general case once we specify the underlying 
cosmological model.

The method relies on a prediction of  $v$ from the observed distribution of 
galaxies in the survey. Since $r=cz-v$ this prediction allows an estimate of the 
 {true} absolute magnitude, 
\begin{equation}
M=m-15 -5\log r =  M_0 - 5\log(1-v/cz)
\label{eq:shift}
\end{equation} 
where the { measurable} absolute
 magnitude $M_0=m-15-5 \log cz$ is determined from observations.
Because the peculiar velocity of a galaxy  is uncorrelated with its true absolute magnitude, 
constraints on the underlying   velocity field can be 
derived  by demanding that 
the distribution of the  magnitudes, $M$,   is consistent with  a reference 
distribution function (i.e.   luminosity function). 
 The equations of gravitational instability theory relate 
the underlying mass density contrast, $\delta_m$, to the peculiar velocity field $\bf u$. 
Further, we will use the linearized equations in which the 
 relation solely depends on 
$f(\Omega)$.
For  linear biasing, $\delta_g=b\delta_m$,  between mass and galaxies 
the appearance of $f$ is replaced by the  single parameter $\beta=f/b$. Therefore, 
the method presented here will focus on constraints on $\beta$ only. More sophisticated 
models for the velocity field involving additional parameters will not be discussed here.
In principle,  the constraint  can be obtained without  resorting to the luminosity function 
simply by minimizing the variance of  $M_0-5\log(1-v(\beta)/cz)$ with respect to $\beta$. However, 
much tighter  constraints are obtained from the full distribution.
We define the luminosity function, $\Phi(M) $,  expressed in terms of the absolute magnitudes, 
as the number density of galaxies per unit magnitude. The probability, $P(M_0|cz, v)$, of observing a galaxy 
having an observed magnitude $M_0$ in a flux limited sample, depends on
its redshift, $cz$, and its radial peculiar velocity, $v$, and is well approximated as  \citep{nbd11a}
\begin{equation}
\label{eq:pmzc}
P(M_0|cz, v)=\frac{\Phi(M)}{\int_{-\infty}^{M_l} \Phi(M) dM      }\; ,
\end{equation}
where  $M_l=M_{0l} -5\log(1-v/cz)$ and $M_{0l}=m_l-15-5\log cz$.
The expression is valid as long as the relative errors on the measured redshift are small
($\sigma_{cz}/cz\ll1$)
The  probability distribution of the whole sample of galaxies is the product of the single probabilities
\begin{equation}
P_s=\Pi_i P(M_{0i}|cz_i,v_i)\; ,
\label{eq:psum}
\end{equation}
where $i$ runs over all galaxies of the sample.
Given a form  for $\Phi(M)$,
 the parameter $\beta$ is then constrained by 
maximizing $P_s$ in which the dependence on $\beta$ is via the $v(\beta)$ as inferred from the spatial 
distribution of galaxies.
{In principle one could use the ``nonparamatric" fit  methods \citep{efs88,DH82} 
to approximate $\Phi(M)$. However, here we will only apply the method  to the 2MRS sample which 
is reasonably approximated by a Schechter luminosity function \citep{w09}. }
Therefore, we assume  $\Phi(M) $ is well approximated by
a Schechter form  \citep{schechter}
\begin{eqnarray}
\label{eq:shform}
\nonumber
\Phi(M)&=&0.4\ln(10) \Phi^* 10^{0.4(\alpha+1)(M_*-M)}\\
&\times &{\rm exp}\left(-10^{0.4(M_*-M)}\right)\;.
\end{eqnarray}
The normalization  $\Phi_*$ does not concern us here.
The shape parameters  $M_*$ and $\alpha$ generally  depend on the 
 galaxies'  type,  { redshift} and band of observation. 
 In terms of the luminosity ($ M=-2.5\log L +const$), this function acquires the simpler form 
\begin{equation}
\label{eq:shformM}
\Phi(M(L))=0.4\ln(10) \Phi^* \left( \frac{L}{L_*}\right)^{1+\alpha}
{\rm exp}\left(-\frac{L}{L_*}\right)\; .
\end{equation}
Inserting all this expression into  (\ref{eq:pmzc}) gives
\begin{equation}
\label{eq:papp}
P(M_0|cz; v)=\frac{0.4 \ln (10) \left(\frac{L}{L_*}\right)^{1+\alpha}{\rm e}^ {-L/L_*}}{\Gamma\left(1+\alpha,L_l/L_*\right)}
\end{equation}
where $L/L_*=(1-v/cz)^210^{0.4(M_*-M_0)}$, $L_l/L_*=(1-v/cz)^210.^{0.4(M_*-M_{0l})}$.

To summarize, given the observed absolute magnitudes $M_0$ for a sample of objects, a Schechter model
for the luminosity function and linear theory prediction for 
the underlying velocity field $v(\beta)$, we determine $\beta$, $\alpha$ and $M_*$ by minimizing 
Eq.~\ref{eq:psum}.

\section{General assessments of the method}
There are three sources of error  which affect 
the derivation of $\beta$ in the method  presented here. 
a) ``shot-noise" error resulting from  the finite number  galaxies, 
b) cosmic variance due to variations in the large scale structure in volumes 
of the universe comparable in size to the volume probed by the redshift  survey at hand
and c) inaccuracies in the peculiar velocity reconstruction.
In this  section we offer a general assessment of the applicability of the method to data at higher redshifts.
To compensate for the degrading of the signal  ($\sim v/cz$) with redshift, surveys with large number of galaxies need to be invoked. 
We will only consider here shot-noise error.
Peculiar velocity reconstruction errors and 
cosmic variance (of surveys with similar volumes) 
 scale  with redshift in a similar manner to the signal  and hence their corresponding relative error
  will not depend on redshift. 
  
To the  list of errors above we do not add biases introduced by adopting a specific form for the luminosity function 
since, as we will argue in \S\ref{sec:sas}, these biases are expected to be insignificant. Further, 
the issue  becomes completely irrelevant for future surveys 
where the  number of galaxies is large enough to allow the use 
of ``nonparametric fit" techniques for modeling the luminosity function, relaxing the need for  specific parametric 
forms.

\subsection{Sensitivity to ``shot-noise" and redshift}

\label{sec:sno}

Assume a cubic\footnote{Any survey geometry with boundary conditions allowing 
robust velocity reconstruction will suffice for our purposes.} region at high redshift $cz$ containing  a flux limited  sample of $N$ galaxies with redshifts $cz_i\approx cz$.
From the galaxy distribution in this region, derive the peculiar velocity field as a function of $\beta$.
Consider  a Schechter form for the luminosity function and assume $cz$ is large enough that  the limiting luminosity $L_{l}$ of a galaxy that could be 
observed is significantly larger than $L_*$.  In the limit $L_l\gg L_*$,  the $\Gamma$  function 
in the expression for $P(M_0|cz; v)$ in
(\ref{eq:papp})
  is well approximated as 
$\Gamma(1+\alpha,L_l/L_*)=(L_l/L_*)^\alpha \exp(-L_l/L_*)$ so that, 
\begin{equation}
P(M_0|cz; v)=0.4 \ln (10) \frac{L}{L_*}
\left(\frac{L}{L_l}\right)^\alpha
{\rm e}^ {-(L-L_l)/L_*}.
\end{equation}
This approximation will allow an  analytic expression for  $\beta$ by 
minimizing $-\ln P_s=-\sum \ln P(M_{0i}|cz_i,v(\beta))$.
For simplicity, we further approximate $v_i(\beta)=F(\beta) v_{1i}$ where $v_{1}$ is the line of sight velocity  
reconstructed with $\beta=1$. For linear velocity reconstruction from the galaxy distribution in 
real space $F(\beta)=\beta$. But for reconstruction from redshift space data
a significantly better scaling is $F=2.5\beta/(1+1.5\beta)$ \citep{DN10}. 
This is not an exact result,  but it suffices here since we are only 
interested in a general assessment of the expected error on $\beta$.
Since $\beta $  appears only   via $F(\beta)$, we 
will perform the minimization with respect to $F$ and write the result in terms of $\beta$ at the 
end of the calculation. 
 Using $L=(1-F v_{1i}/cz_i)^2 L_0 $ and $L_l=(1-Fv_{1i}/cz_i)^2 L_{0l}$
we get 
\begin{eqnarray}
&&\frac{\partial \ln P_s}{\partial F}
=\\
&& \frac{\partial }{\partial F}
\sum\left[ \ln\left(1-F\frac{v_{1i}}{cz_i}\right)^2 -\frac{L_{0i}-L_{0li}}{L_*}\left(1-F \frac{v_{1i}}{cz_i}\right)^2\right]\nonumber \\ \nonumber
&&=2\sum\left[ \frac{\Delta L_i}{L_*} \frac{v_{1i}}{cz_i}-F \left(\frac{v_{1i}}{cz_i}\right)^2\left(1+\frac{\Delta L_i}{L_*}\right)\right] \; , \nonumber 
\end{eqnarray}
where $\Delta L_i=L_{0i}-L_{0li}$ and in the last step we have neglected   $O(v_1/cz)^2$  terms
and  assumed that the Hubble flow-like $\sum v_{1i}/cz_i$ is negligible compared to the other terms. 
The $1\sigma$ shot-noise error on $F$  is, therefore,
\begin{eqnarray}
\label{eq:dF}
\delta F& =& \left(\frac{-2}{\partial^2 \ln P/\partial F^2}\right)^{1/2}\\
& =&
\left[\sum\left(\frac{v_{1i}}{cz_i}\right)^2\left(1+\frac{\Delta L_i}{L_*}\right)\right]^{-1/2} \nonumber \; .
\end{eqnarray}
This expression can be easily estimated when the luminosity, $L_{0}$, is computed from 
the actual distances, i.e.
$cz_i=r_i$. This means that $L_{0}$ is the true intrinsic luminosities and, therefore, 
 $\Delta L$ and $v_1$ are uncorrelated.   Further,  in the limit $L_l\gg L_*$,  
the average $<\Delta L/L_*> $ is unity.
From  $F=2.5\beta/(1+1.5\beta)$ we get, 
$\delta \beta \approx 0.4 (1+1.5\beta)  \delta F$. With all this, equation (\ref{eq:dF}) gives 
\begin{eqnarray}
\label{eq:betaerr}
\delta \beta &=&0.4(1+1.5\beta) (2N)^{-1/2}\frac{c\tilde z}{\sigma_{v_1}}\\
&=& 0.044(1+1.5\beta) \left(\frac{10^5}{N}\right)^{1/2} \frac{\tilde z}{0.1}\frac{600}{\sigma_{v1}}\; ,\nonumber 
\end{eqnarray}
where $\tilde z=<1/z^2>^{-1/2}$,and $\sigma_{v1} $ is the rms value of $v_1$  in $\kms$. 
The analysis of 
\cite{DN10} gives, on large scales,  $\sigma_{v1}\sim 600\kms$ for $\beta=1$, at $z=0$.
 
{  The scaling $\delta \beta \propto N^{-1/2} \tilde z$  in Eq.~\ref{eq:betaerr} should be approximately 
valid also when the condition $L_l\gg L_*$
is not strictly satisfied.  
For 2MRS  this scaling  gives roughly the same error on $\beta$ using distant galaxies 
with $cz>4000\kms$ as using galaxies with the lower redshifts }.

\subsection{Sensitivity to assumed form of the  luminosity function}
\label{sec:sas}
\def\vo{\frac{v_{1i}}{cz_i}}
\def\vos{{v_{1i}}/{cz_i}}
 
It is instructive to assess  the sensitivity of  $\beta$ on 
the assumed form of the luminosity function.
 We give here a simple example 
in which the assumed luminosity function differs greatly from the true form, yet the estimate for $\beta$
is unbiased.  Consider an ideal  volume-limited sample of galaxies with a true luminosity function of a Schechter form with 
 $\alpha=0$, i.e. an exponential distribution.  Let is try to recover $\beta$ assuming a gaussian form for the luminosity function, $P(M_0|cz,v)\propto \sigma_L^{-1}(1-F v_1)^2 {\rm e}^{-(L -L_m)^2/2\sigma_L^2}$, where, like in the previous section, 
$v(\beta)=F(\beta) v_1$.
Minimizing  the quantity, $-\sum \ln P(M_{0i}|cz_i,F v_{1i}) $,  with respect to $L_m$, $\sigma_L$ and $F$, yields, respectively,
\begin{eqnarray}
0&=&L_m-<L>\; ,\\
0&=& \sigma_L^2-<(L -L_m)^2> \; , \nonumber \\
0&=&\sum_i\left[\frac{\vo}{1-F\vo}-\frac{L_0(L_i-L_m)}{\sigma_L^2}\vo(1-F\vo)^3\right]. \nonumber 
\end{eqnarray}
 For simplicity we further assume that $L_0$ is computed with $cz_i=r_i$ so that 
the true solution $\beta=0$ (the generalization to cases where the true $\beta$ is different from zero is trivial).  Hence, $L_0$  are equal to the true luminosities and, therefore,  follow an exponential distribution
for which $ <L_{0}>=L_*$ and $<L_0^2>=2L_*^2$. 
Further, 
 there is no correlation between $L_0 $ and $\vo$, meaning  that the average of products of powers of $(1-F\vos)$ and $L_0$ is
the product of the averages. In the limit  $N\rightarrow \infty$, straightforward algebraic manipulation then yields 
$F(\beta)=0$, $L_m=L_*$, and $\sigma_L^2=L_*^2$. 
Therefore, in this example, where the assumed and true luminosity functions differ grossly, the 
best fit  $\beta$ is unbiased.
This is not surprising since the underlying principle of the method is a reduction in the spread of $L$. Therefore, the  assumed 
form of the   luminosity function should only  affect the weighting given to galaxies in a  certain luminosity range rather, than the 
best fit $\beta$. 
 Assuming a wrong luminosity function increases the random error
on $\beta$ but does not introduce any systematic biases.

\section{Application to 2MRS}

In this section we apply the method outlined above to the all sky 2MRS
consisting of  23,200 galaxies down to the 
magnitude  $K=11.25$. 
Details about the catalog, including the precise completeness, sky coverage and selection effects can be found in
 \citep{fhuch}. The preparation of the catalog for the purpose of the application of the method is 
is done similarly to \cite{DN10}.
The peculiar velocity field is derived from the galaxy distribution in the  2MRS 
for an array of $\beta$ values using  the linear theory  methodology of \cite{ND94}.
The derived velocity field is robust within $cz<10^4\kms$, above that redshift 
discreteness effects become important. Hence we limit the analysis to the $~18,000$ galaxies with $cz<10^4\kms$.
In the derivation of the velocity fields, the galaxy distribution is smoothed with a gaussian window 
of constant width of $400\kms$. To further remove strong nonlinearities the 
derived three-dimensional velocity fields are smoothed with an Gaussian window of constant width,
$R_s$.
Linear theory  recovers the flow pattern  reasonably  well  even at  $\delta \ltsim 1$, but not 
beyond \citep{nussetal91,BEN02}. 
Therefore,  although  
 the peculiar velocity field is predicted from the distribution of all galaxies,
in the maximization of $P_s$ to assess the 
robustness of the method
we remove  galaxies in
regions with density contrast  higher than $\delta_{cut}$  as listed in Table~\ref{tab:s}
for both values of $R_s$. 
Using the expression (\ref{eq:papp}) we  minimize
$-\ln P_s=  -\sum_i \ln P(M_{0i}|cz_i; v(\beta))$ (the summation is over all galaxies) with respect to
 $\beta$ and the Schechter parameters, $\alpha$ and $M_*$.
 
 \subsection{Error estimation based on mock 2MRS catalogs}
 \label{se:mock}
The overall expected  errors in $\beta$, including possible biases,  are based on
mock catalogs designed to 
match the general properties of the 2MRS.  For this purpose we use 135 2MRS mock catalogs very similar to those compiled by 
\cite{DN10}. 
These catalogs are extracted from a parent mock catalog of the Two Micron All Sky Survey (2MASS)
\citep{skrut}.
The parent catalog is generated from  the Millennium simulation \citep{mill} using semi-analytic 
galaxy formation models \citep{delucia}.   All 135 2MRS catalogs satisfy the following conditions. 
a)  The central ``observer" in each mock is selected to reside in a galaxy with 
a quiet velocity field within $500\kms$,  similar to the observed universe,
b)  The motion of the central galaxy is 500 to 700 $\kms$, and 
c)  The density in the environment of the local group, averaged over a sphere of 400 km/s radius ,  is  less than twice the normal. These conditions select central observers that are similar to the conditions of our own Local Group.

The luminosity function in  each mock is  approximated by a Schechter form. 
Using the galaxy distribution in each mock, the corresponding peculiar velocity field 
is generated for an array of  $\beta$ values. 
The mean and standard deviation of those best values from the 135 catalogs is 
 $\beta=0.49 \pm 0.13$. 
 The cosmological density parameters of 
the Millennium simulation are 
$\Omega=0.25$ and $\Lambda=0.75$ for matter and cosmological constant. This 
yields $f=\Omega^{0.55}=0.47$ \citep{Lind05}.  A calculation of the rms galaxy density fluctuations 
of the mocks yields $b=1\pm 0.1$ for the mocks. 
The $1\sigma $ rms scatter 
of the best fit  $\beta $ values is $0.13$. This scatter reflects ``shot-noise" errors due to the 
finite number of galaxies, 
cosmic variance due to the limited 
volume covered by the 2MRS, and  inaccuracies in the reconstruction of the 
peculiar velocity field by means of linear theory. 
 Shot-noise is subdominant when the 
method is applied to all  2MRS galaxies within $10^4\kms$. Since $P_s$ contains no information about cosmic variance and 
reconstruction errors,  the  
width of $-2\ln P_s$ versus $\beta$, reflects shot-noise errors only. 
From the width of $-2\ln P_s$ we get $1\sigma$ shot-noise error of $\delta \beta_{sn}\approx 0.055 $ in the application 
to 2MRS within $cz<10^4\kms$.
Cosmic variance error could be estimated by running the method with the actual 
velocity field of galaxies in the mock. This amounts to $\delta \beta_{cv}\approx 0.09$.
Adding errors in quadratures, we infer a velocity reconstruction error of $\delta \beta_{vrec} \approx
0.1$, comparable to $\delta \beta_{cv}$.

We will apply the method to various cuts of the 2MRS.
All corresponding errors listed  in Table~\ref{tab:s} are based on similar cuts taken from the mocks. 
Shot-noise errors,  $\delta \beta_{sn}$, will be  treated as (nearly) independent of $\beta$ (see \S\ref{sec:sno}), while 
cosmic variance, $\delta \beta_{cv}$ and velocity reconstruction errors, $\delta \beta_{vrec}$, are assumed to be proportional to $\beta$, 
because of the $\beta$ dependence in the reconstructed field.

 \subsection{$\beta$ from the real 2MRS}
 
\begin{figure} 
\centering
\epsscale{1.1}
\plotone{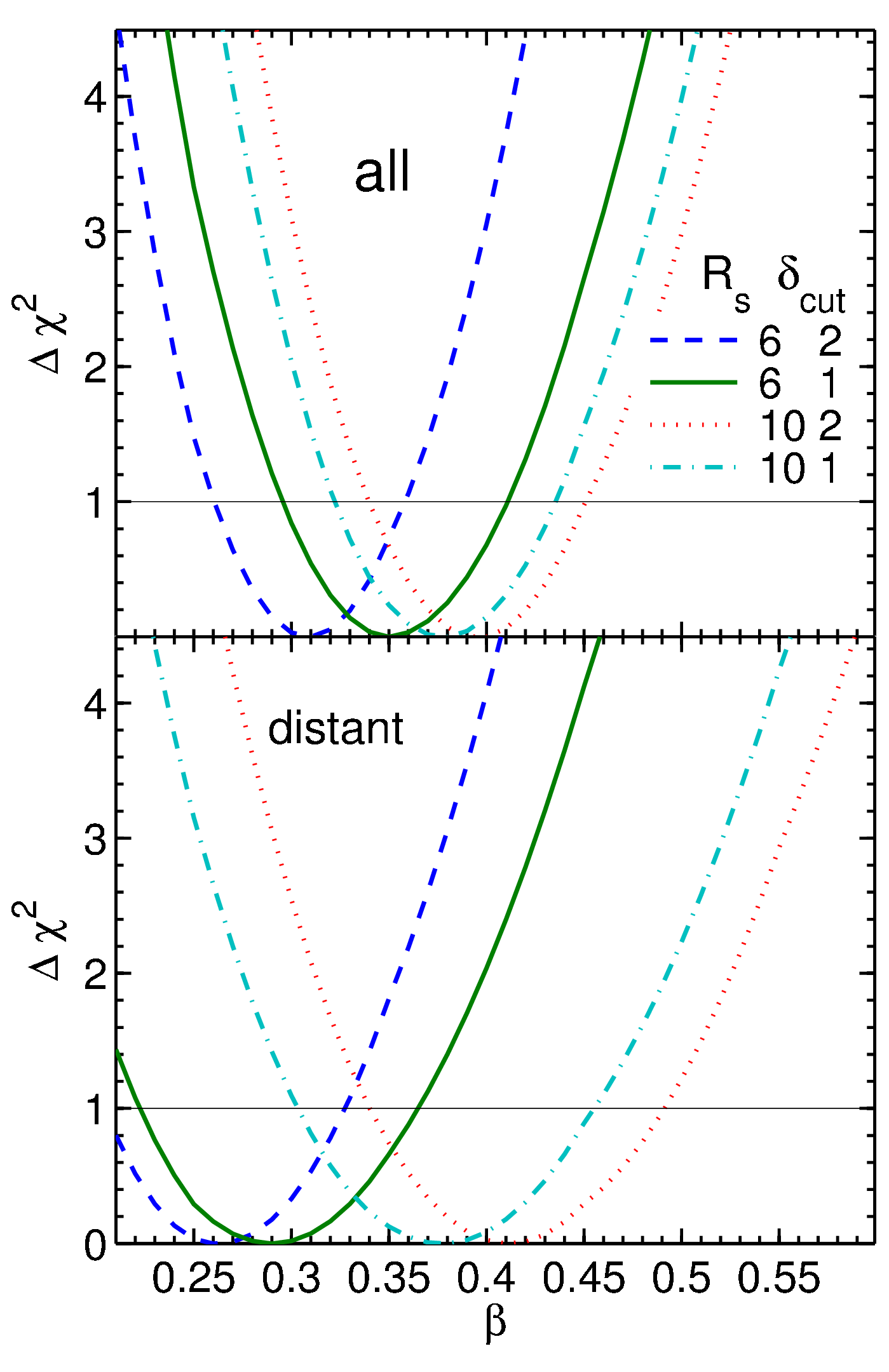}
\vspace{0pt}
\caption{
 {\it Top:} Curves of $\Delta \chi^2$ as a function of $\beta$ for $R_s=6\hmpc$ and $10\hmpc$, and 
$\delta_{cut}=2$ and 1, as indicated in the figure. The width of the curves reflects uncertainties 
due to the finite number of galaxies (``shot-noise") and do not include cosmic variance and inaccuracies in the 
peculiar velocity reconstruction. 
{\it Bottom:} The same as the {\it top} panel, but  excluding  nearby galaxies with $cz<4000\kms$
in the minimization of $P_s$. The widths of curves here are substantially larger than 
in  the {\it top} panel
due to the smaller number of galaxies and the larger redshifts. 
}
\label{fig:chi}
\end{figure}

The results are presented in Table~\ref{tab:s} and 
the top panel of Fig.~\ref{fig:chi}. The plotted quantity, $\Delta \chi^2$, is 
$-2\ln P_s$ evaluated as a function of $\beta$ minus its value at the minimum. 
In the figure, the parameters $(\alpha,M_*)$ are fixed at their best fit values. 
The overall errors listed in Table~\ref{tab:s} are based on the analysis of the mocks in \S\ref{se:mock}.
The $1\sigma$ shot-noise, $\delta_{sn}$,  given by the width of the curves at $\Delta\chi^2=1$
are significantly smaller than the corresponding total errors quoted in the Table~\ref{tab:s}, i.e.
 $\delta \beta_{sn}$  is not the main source of error for the 2MRS catalog.
Table~\ref{tab:s} lists results only for $R_s=6\hmpc$, however,  an application of  the method 
with $R_s=10\hmpc$ and $12\hmpc$ yields  consistent results within the $1\sigma$ errors in the table.
The derived $\beta$  for both choices of $\delta_{cut}$ are consistent with \cite{DN10} and 
the values of $\alpha$ and $M_*$ agree well with \cite{w09}.

It is useful to examine how the best fit $\beta$ changes when in the  maximization of $P_s$
we include only distance galaxies with $cz>4000\kms$.
The results are given in the Table~\ref{tab:s} and in 
the bottom panel of Fig.~\ref{fig:chi} showing $\Delta \chi^2$ versus $\beta$ for this case. 
Shot-noise as indicated by  width of 
curves increases with respect to the full sample  (see top panel of the same figure).
 The constraints from the distant cut are still 
reasonably tight and, within the total $1\sigma$ errors,  are fully consistent with those obtained from  the whole sample.

We also applied the method to  galaxies in the Northern and Southern 
 Galactic hemispheres, separately. 
The results, listed  in the Table~\ref{tab:s}, show  a non-negligible difference between 
the derived $\beta$ in the two hemispheres. 
The corresponding  error on $\beta$ is based on the application of the method to 
 ``Northern" and ``Southern" hemispheres in the mocks.
  We attribute the difference between the North and the South to 
cosmic variance and confirm that it is consistent with the mocks.
 For each mock catalog we compute $\beta_{north} $
and $\beta_{south} $ and find an rms value
 $<(\beta_{_{\rm north}}-\beta_{_{\rm south}})^2>^{1/2}=0.2$.

\begin{table*}
\begin{center}
\vspace{0pt}
\caption{Derived parameters, $\beta=f/b$, $\alpha$ and $M_*$ for various
cuts of the 2MRS. 
Galaxies in regions with density contrast larger than $\delta_{cut} $ and redshifts
less than $cz_{cut}$ are excluded from the maximization of $P_s$.
The quoted errors on $\beta$ are based on mock catalogs and include shot-noise, 
cosmic variance and inaccuracies in the peculiar velocity reconstruction.
The results given here correspond  to $R_s=6\hmpc$. }

\vspace{10pt}
\label{tab:s}
\begin{tabular}{|c|c|c|c|c|c|}
\hline \hline
sky &  $cz_{cut}$  & $\delta_{cut}$  & fraction & $(\alpha, M_*)$  & $\beta\;  \pm 1\sigma$ error \\
      &  $\kms$                                            &                         &        &                         &     \\ 

\hline 
all &  0 & $\infty $ &  $100\% $  & $(-0.92, -23.14)$ & $0.30\pm 0.10$  \\  
all &  0 & $2 $ &  $79\% $  & $(-0.90, -23.19)$ & $0.31\pm 0.10$  \\  
all &  0 & $1 $ &$58\%$ & $(-0.89, -23.14)$ & $0.35\pm 0.10$ \\  
all & 4000 & $2 $& $ 61\% $ & $(-0.84, -23.15)$ & $0.26\pm 0.15$ \\  
all & 4000 & $1 $ &$44\%$  & $(-0.81, -23.09)$  &$0.29\pm 0.15$ \\

north &  0 & $2 $ &$ 41 \%$ & $(-0.88, -23.16)$  & $0.23\pm 0.10$ \\
north & 0 & $1 $ &$30 \%$ & $(-0.85, -23.09)$ & $0.26\pm 0.11$   \\  

south & 0 & $2 $ &$38 \%$ & $(-0.93, -23.20)$ & $0.40\pm 0.17$ \\ 
south & 0 & $1 $ &$29 \%$ & $(-0.93, -23.17)$ &  $0.41\pm 0.17$ \\
\hline
\end{tabular}
\end{center}
\end{table*}

\section{Discussion and Conclusions}
We have presented a new method to determine $\beta$
from galaxy redshift surveys.   
The method 
is entirely independent of distance indicators such as the Tully-Fisher relation and 
of analyses of  anisotropic correlation functions, $\xi(r_p,\pi)$ \citep[e.g.][]{k87}. 
As a preliminary application of the method, we have resorted to the  K=11.25 flux limited 2MRS all sky survey.    In the maximization procedure of $P(M_0|cz,v(\beta))$, galaxies 
in dense regions should  be excluded since the flow pattern in these regions is not 
well recovered by linear theory.  For our adopted density contrast cut $\delta_{cut}=1$, 
we get  a best fit  $\beta=0.35\pm 0.1$  for velocities smoothed with a gaussian window of $6\hmpc$ in width. 
However, the results obtained with $\delta_{cut}=2$ and $\delta_{cut}=\infty$ are consistent with this 
best fit value  and are reported in the Table~\ref{tab:s}.
These constraints on $\beta$  agree very well  with   those  of
\cite{DN10}  who compared the  peculiar velocities of the Spiral Field I Band (SFI++)  catalog of spiral
galaxies \citep{mas06, spring07} with the velocity field predicted from the 2MRS.

There are three sources of uncertainties which contribute to the error budget on the 
estimated $\beta$:  
\begin{description}
\item[(1)] ``shot-noise"  due to the finite number of galaxies  
\item[(2)]  cosmic variance which reflects the variation
 of the large scale structure in random volumes comparable in size to 
the volume covered by the data set under consideration. 
\item[(3)]  inaccuracies in the linear methodology 
 for reconstructing the peculiar velocity from the galaxy distribution. 
 \end{description}
 
For the 2MRS sample of $\sim 18,000$ galaxies within $cz=10^4\kms$, shot-noise is subdominant. 
Increasing the number of galaxies in the sample without probing larger volumes will not tighten the constraint significantly. 
Cosmic variance can only be reduced by pushing toward deeper and larger surveys.
The main galaxy sample of the
 Sloan Digital Sky Survey  (SDSS) \citep{2002AJ....124.1810S}
 already offers this opportunity.
Let us consider the $\sim 7500$ deg$^2$
patch around the Northern Galactic cap
in the SDSS-DR7 release \citep{2009ApJS..182..543A}.
A shell with $\Delta z=0.1$  centered
at $z\sim 0.135$, close to the peak of the galaxy $dN/dz$
has a comoving volume $\sim 20$ times larger than that of the 2MRS
and it could be divided into $\sim 12$ independent cubes of $200\hmpc$,
each   containing as many galaxy as 2MRS.
Applying our method to each of them would dramatically decrease 
cosmic variance and errors in the reconstructed  velocities.
According to Eq.~\ref{eq:betaerr} the now dominant shot-noise error 
would be $\delta\beta \sim 0.06$, twice as small as in the 2MRS case,
as shown in Table~\ref{tab:dbeta}.

Shot-noise errors increase linearly with redshift and they will be the limiting 
factor in limiting the precision of constraining $\beta$ from the method presented in this paper.
Next-generation, large redshift surveys like Boss, BigBoss and EUCLID
will allow an application of  our method  to measure $\beta$ out to $z\sim 1$. 
We have applied Eq.~\ref{eq:betaerr} to estimate the expected shot noise error on $\beta$ 
for all these surveys. The results are listed in Table~\ref{tab:dbeta}. 
The ongoing SDSS-III (BOSS) survey
will target highly luminous galaxies with a nearly constant 
number density  $n \sim 3\cdot 10^{-4} h^3$ Mpc$^{-3}$ \citep{2011arXiv1101.1529E}
over the redshift range  $[0.2-0.6]$. Because of the relatively low number density of objects
our method will measure $\beta$ with uncertainties twice as large as for the 2MRS case. 
The  BigBOSS survey will expand Boss, both in sky coverage and depth.
The expected number of  ELGs 
 \citep{2011arXiv1106.1706S} will be large enough to decrease 
 the  errors on $\beta$ significantly. In fact, the reduction of the relative 
 errors will be even larger since, in this redshift rage, $\beta$ is an increasing function of $z$.
 Finally, the EUCLID survey \citep{2009arXiv0912.0914L}   will 
 reduce  errors even further in the redshift range $z=[0.7-1.0]$.

How do the $\beta$ estimates obtained with our method compare with those
obtained from the analysis of the anisotropy pattern of the two-point correlation function
$\xi(r_p,\pi)$ ? At low  ($z  < 0.2$) redshift the $\xi(r_p,\pi)$  method has been applied
to SDSS  \citep{2006PhRvD..74l3507T}
and 2dF  \citep{2003MNRAS.346...78H} allowing to estimate 
$\beta$ with an error $\delta \beta =0.1-0.15$. 
The application of our method to 2MRS already gives $\beta$ with
a similar precision and the upcoming application to SDSS-II data \citep{bernardi2003} will allow us to 
estimate  $\beta$ with a precision sufficient to test the validity of 
popular alternative gravity model, like the 5-dimensional brane-world of
 \cite{DGP,Wei}  as well as the possibility of a coupling between the
 dark energy and the dark matter sectors \citep{2008PhRvD..77h3508D}.
 
The first measurement of the growth rate at larger redshifts 
has been performed by \cite{2008Natur.451..541G}. 
From the observed $\xi(r_p,\pi)$ of VVDS galaxies they obtained
$\beta(z=0.77)=0.70 \pm 0.26$. 
More recently, the Wiggle-z experiment  \citep{2011MNRAS.tmp..834B}
has measured the normalized growth rate $f(\Omega)\sigma_8$  
in 4 redshift bins to $z=0.9$ with an error $\delta f(\Omega)\sigma_8 \sim 0.1$.
This is already comparable with the expected performance of our method
on future datasets. 
Indeed, our method does not compare favorably with the analysis of 
galaxy clustering in future surveys. In the case of EUCLID, the goal is to
estimate the growth rate from $\xi(r_p,\pi)$ 
with a precision of $0.01$ at $z=1$, if 
the rms mass fluctuation $\sigma_8$ can be determined accurately
\citep{2009JCAP...10..004S}.
In this case our 
alternative method to measure $\beta$
will constitute an effective way to keep systematic errors 
below  $\delta \beta = 0.1$. 
It is important to note that  if galaxy bias will only be constrained at
the $10 \%$ level then both methods, the analysis of $\xi(r_p,\pi)$ and the one
proposed here,  will constrain the   growth rate $f(\Omega)$ with similar precision.
Further, on-going and planned redshift surveys will deliver velocity dispersion for all the elliptical  galaxies will be obtained.   Using the 
Faber-Jackson relation \citep{fj76}, between luminosity and velocity dispersion of   elliptical galaxies, in conjunction with our method will produce 
even tighter constraints. Additional constraints on $f$ could also be obtained from 
the expected large scale supernova survey \citep[e.g.][]{newman}.

\begin{table*}
\begin{center}
\vspace{0pt}
\caption{Expected shot-noise errors on $\beta$ estimated from 
Eq.~\ref{eq:betaerr} for present and future galaxy redshift surveys.
To minimize evolutionary effects all errors are estimated in 
 bins $\Delta z=1$. The reference $\beta(z)$ function 
 has been obtained by considering the growth rate of 
 a $\Lambda$CDM model of \cite{Lind05} and a linear bias
 $b(z)=\sqrt(1+z)$ as in \cite{2011arXiv1101.2453D}.
The references to the surveys' parameters are given in the text.}

\vspace{10pt}
\label{tab:dbeta}
\begin{tabular}{|c|c|c|c|c|}
\hline \hline
Survey & sky &  $\langle z \rangle $  & $N_{\rm gals.}$   & $\delta \beta\;$  \\
              & $deg^2$  &                          &        &                           \\ 

\hline 
SDSS-II & 7500 &  0.135 & $3\cdot 10^5$ & 0.07 \\
Boss & 10000 &  0.25 & $9\cdot 10^4$ & 0.18 \\
Boss & 10000 &  0.35 & $3\cdot 10^5$ & 0.17 \\
BigBoss & 14000 &  0.45 & $5\cdot 10^5$ & 0.16 \\
BigBoss & 14000 &  0.55 & $1.3\cdot 10^6$ & 0.14 \\
BigBoss & 14000 &  0.65 & $2\cdot 10^6$ & 0.13 \\
EUCLID & 15000 &  0.70 & $3.5\cdot 10^6$ & 0.11 \\
EUCLID & 15000 &  0.80 & $6.5\cdot 10^6$ & 0.09\\
EUCLID & 15000 &  0.90 & $7\cdot 10^6$ & 0.09 \\
EUCLID & 15000 &  1.00 & $7.5\cdot 10^6$ & 0.10 \\
EUCLID & 15000 &  1.10 & $7\cdot 10^6$ & 0.11 \\

\hline
\end{tabular}
\end{center}
\end{table*}

Quasi-linear (i.e. the mildly non-linear regime) dynamical reconstruction methods 
 offer a substantial improvement in the accuracy 
of the predicted peculiar velocities.
In particular, the Fast Action Method \citep{NB00,be02} 
adaptation of Peebles' least action principle  \citep{sh95}  is fast and easy to implement. 
This method is significantly better than linear theory for the reconstruction of the peculiar velocity field on small scales
and also in dense regions. 
The linear theory relation between mass and velocity solely depends on the growth rate
$f(\Omega)=\Omega^\gamma$.  
In the quasi-linear regime, there is an additional explicit dependence on $\Omega$, raising the possibility 
of separate constraints on  $\Omega$ and the growth index $\gamma$. 
 However, this does not seem a promising route to break the $\Omega-\gamma$ degeneracy since the explicit $\Omega$ dependence is very weak ($\sim \Omega^{0.2}$)  \citep{NC98}. 

Another improvement  could be achieved 
by use of nonparametric fit  techniques for modeling the galaxy luminosity distribution. 
Although the Schechter form is a good approximation for the 2MRS, it is likely less successful for 
larger and deeper data sets.  A variety of such nonparametric fit methods exist  \citep[e.g.][]{efs88,DH82}  could easily be incorporated in the method presented here.

Angular coherent   photometric mis-calibrations  are 
likely to contaminate the data at some level. However, as long as the  density field 
inferred from the galaxy distribution is not significantly affected, the method should yield an unbiased $\beta$. This is 
because the underlying velocity field should be uncorrelated with observational mis-calibrations. If systematic biases introduce serious spurious 
modes in the density field then it is doubtful if the data could be useful 
for any analysis of large scale clustering.

\section{Acknowledgments}
We thank Gerard Lemson  for help in generating the mock 2MRS catalogs. 
This work was supported by THE ISRAEL SCIENCE FOUNDATION (grant No.203/09), the German-Israeli Foundation for 
Research and Development,  the Asher Space Research
Institute and  by the  WINNIPEG  RESEARCH FUND.
MD acknowledges the support provided by the NSF grant  AST-0807630. 
EB  acknowledges the support provided by
MIUR PRIN 2008 'Dark energy and cosmology with large galaxy surveys' and by
Agenzia Spaziale Italiana (ASI-Uni Bologna-Astronomy Dept. 'Euclid-NIS' I/039/10/0)
EB also thanks the Physics Department and the Asher Space Science Institute-Technion for the kind hospitality.

\bibliography{gbeta.bib}

\end{document}